\documentclass[pra,twocolumn,
superscriptaddress,
showpacs,
aps
]{revtex4-1}
\usepackage{graphicx}
\usepackage{amsmath}
\usepackage{amsfonts}
\usepackage{amssymb}
\usepackage{hyperref}
\usepackage{color}

\newcommand{\beq}{\begin{equation}}
	\newcommand{\eeq}{\end{equation}}

\usepackage{dsfont}
\usepackage{placeins} 
\usepackage{bm}           

\begin{document}

	\title{Regular and bistable steady-state superradiant phases of an atomic beam traversing an optical cavity}
	\author{Simon~B.~J\"ager}
	\affiliation{JILA and Department of Physics, University of Colorado, Boulder, Colorado 80309-0440, USA.}
	\author{Haonan~Liu}
	\affiliation{JILA and Department of Physics, University of Colorado, Boulder, Colorado 80309-0440, USA.}
	\author{Athreya~Shankar}
	\affiliation{JILA and Department of Physics, University of Colorado, Boulder, Colorado 80309-0440, USA.}
	\author{John~Cooper} 
	\affiliation{JILA and Department of Physics, University of Colorado, Boulder, Colorado 80309-0440, USA.}
	\author{Murray~J.~Holland}
	\affiliation{JILA and Department of Physics, University of Colorado, Boulder, Colorado 80309-0440, USA.}
	\begin{abstract}
		We investigate the different photon emission regimes created by a preexcited and collimated atomic beam passing through a single mode of an optical cavity. In the regime where the cavity degrees of freedom can be adiabatically eliminated, we find that the atoms undergo superradiant emission when the collective linewidth exceeds the transit-time broadening. We analyze the case where the atomic beam direction is slanted with respect to the cavity axis. For this situation, we find that a phase of continuous light emission similar to steady-state superradiance is established providing the tilt of the atomic beam is sufficiently small. However, if the atoms travel more than half a wavelength along the cavity axis during one transit time we predict a dynamical phase transition to a new bistable superradiant regime. In this phase the atoms undergo collective spontaneous emission with a frequency that can be either blue or red detuned from the free-space atomic resonance. We analyze the different superradiant regimes and the quantum critical crossover boundaries. In particular we find the spectrum of the emitted light and show that the linewidth exhibits features of a critical scaling close to the phase boundaries.
	\end{abstract}
	\maketitle
	\section{Introduction}
	Coupling quantum particles to bosonic modes enables the building of versatile platforms to study driven-dissipative dynamics in various physical setups. Prominent examples include trapped ions~\cite{Burd:2020}, color centers in diamonds~\cite{Angerer:2018}, semiconductor systems~\cite{Rodriguez:2016}, and atoms in optical cavities~\cite{Muniz:2020}. The bosonic modes typically serve as common and intrinsically lossy channels that enable strong interactions. In particular atomic ensembles in optical cavities have been used to investigate many-body effects that are of elementary and fundamental interest, such as exotic quantum phases~\cite{Nagy:2008, Larson:2008, Baumann:2010, Habibian:2013, Landig:2016, Leonard:2017:1, Leonard:2017:2, Kroeze:2018, Landini:2018, Dogra:2019} and collective dissipative dynamics \cite{Domokos:2002, Black:2003, Ritsch:2013, Schutz:2014, Schutz:2016, Xu:2016, Keller:2018}, but are often accompanied by potential technological applications~\cite{Meiser:2009, Schleier-Smith:2010, Bohnet:2012, Pezze:2018, Lewis-Swan:2018, Shankar:2019:1, Shankar:2019:2}. 
	
	An example of such technology is the steady-state superradiant laser~\cite{Meiser:2009, Bohnet:2012}. This laser works in the regime where the lifetime of cavity photons is orders of magnitude shorter than the lifetime of the coherent dipoles. In this regime, coherences are stored in the atoms and are robust against environmental noise~\cite{Meiser:2009, Bohnet:2012, Meiser:2010:1, Meiser:2010:2, Bohnet:2014, Norcia:2016:1, Norcia:2016:2}. Besides this technological feature, this setup has also been connected to time crystals~\cite{Tucker:2018, Iemini:2018, Gong:2018, Kessler:2019, Zhu:2019}, synchronization~\cite{Mori:1998, Acebron:2005, Xu:2014, Zhu:2015, Weiner:2017}, and dynamical phase transitions~\cite{Barberena:2018, Norcia:2018, Jaeger:2019, Muniz:2020, Jaeger:2020}. The rich dynamics of this system is based on effective interactions between the atoms and requires that the atoms remain in the cavity over long time scales. 
	
\begin{figure}[h!]
		\center
		\includegraphics[width=0.75\linewidth]{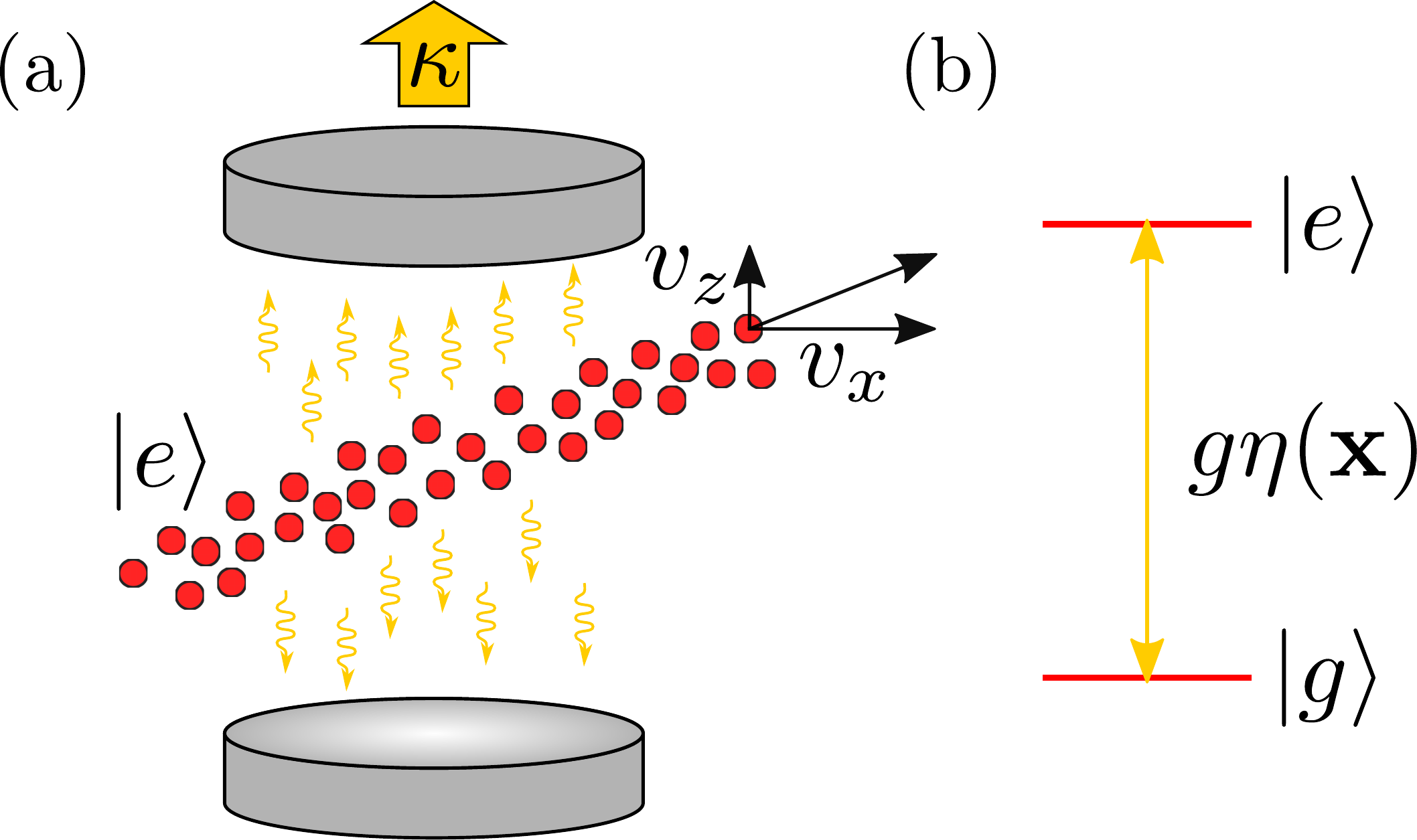}
		\caption{(a) Atoms are preexcited and pass through a lossy optical cavity. (b) Two-level atoms resonantly exchange photons with the cavity mode with a spatially dependent coupling~$g\eta({\bf x})$.\label{Fig:1}}
	\end{figure}

	In this paper, we will investigate superradiant phases that establish and persist on timescales that are much longer than the lifetime of any individual photon or atom in the cavity. In order to show this, we consider an atomic beam  that traverses an optical cavity (see Fig.~\ref{Fig:1}a). 
	A similar system has been studied in~\cite{Liu:2020} for purposes of realistic quantum metrology applications such as active optical clocks~\cite{Chen:2009} and ultra-narrow linewidth lasing in the field~\cite{Kolkowitz:2016, Takamoto:2020}. The superradiant phases that arise from such systems highlight the ability of many-body states to store coherence on timescales exceeding the lifetime of their constituents. 
	
	The paper is structured as follows. In Sec.~\ref{sec:Model} we introduce a semiclassical treatment to describe the dynamics of the atomic beam. In Sec.~\ref{sec:OSR} we determine the parameter regime where the atomic beam will undergo superradiant emission. In Sec.~\ref{sec:SR} we analyze the two occurring superradiant phases and study in detail the crossover between the two phases. We conclude with a  discussion of the results and their implications in Sec.~\ref{sec:Concl}. The Appendix provides additional details of the calculations presented in the main text.
	
	\section{Model}\label{sec:Model}
	We study the dynamics of a collimated atomic beam that passes through an optical cavity. In our model, the atomic beam is composed of atoms that have the same identical velocity $\textbf{v}=(v_x,v_z)$, where $v_x$ ($v_z$) is the longitudinal (transverse) component perpendicular (parallel) to the cavity axis (see Fig.~\ref{Fig:1}a). Each atom possesses internal degrees of freedom that are described as a two-level system representing an optical dipole with transition frequency~$\omega_a$ between its excited~$|e\rangle$ and ground state~$|g\rangle$. We assume throughout this paper that the atoms are preexcited in $|e\rangle$ before they enter the cavity. Once in the cavity, every atom interacts during its transit time $\tau$ with a single cavity mode with linewidth $\kappa$ and frequency~$\omega_c$ that is on resonance, i.e., $\omega_c=\omega_a$. The atom-cavity coupling is characterized by a vacuum Rabi frequency $g$ at the maximum of the cavity mode function $\eta({\bf x})$ (see Fig.~\ref{Fig:1}b).
	
	\subsection{Parameter regime and quantum mechanical description}
	We investigate the regime where the lifetime of cavity photons is much shorter than the atom transit time, i.e., ${\kappa^{-1}\ll\tau}$, and the Rabi splitting due to the coherent atom-cavity exchange is unresolvable, i.e., $\sqrt{N}g\ll\kappa$, where $N$ is the mean intracavity atom number. In this regime, the field mode mediates an all-to-all interaction between the atoms, and exposes the dipoles to quantum noise that physically arises from the vacuum leaking through the cavity output. Consequently, we can adiabatically eliminate the field variables and describe the dynamics of the atomic degrees of freedom using the following Heisenberg-Langevin equations
	\begin{align}
		\frac{d\hat{\sigma}_j^{-}}{dt}=&\frac{\Gamma_c}{2}\eta({\bf x}_j)\hat{\sigma}_j^{z}\hat{J}^{-}+\hat{\mathcal S}^{-}_j,\label{sigma-}\\
		\frac{d\hat{\sigma}_j^{z}}{dt}=&-\Gamma_c\eta({\bf x}_j)\left(\hat{J}^{+}\hat{\sigma}_j^{-}+\hat{\sigma}_j^{+}\hat{J}^{-}\right)+\hat{\mathcal S}^{z}_j,\label{sigmaz}\\
		\frac{d{\bf x}_j}{dt}=&{\bf v}_j\label{hatx}.
	\end{align}
	These equations are presented in the reference frame rotating with frequency $\omega_a$. Here $j$ labels the individual atoms and $\hat{\sigma}_j^{-}=|g\rangle_j\langle e|_j$, $\hat{\sigma}_j^{+}=\left(\hat{\sigma}_j^{-}\right)^{\dag}$ are the annihilation and creation operators of an electronic excitation and $\hat{\sigma}_j^{z}=|e\rangle_j\langle e|_j-|g\rangle_j\langle g|_j$ for atom $j$. The internal degrees together with the position ${{\bf x}_j=(x_j,z_j)}$ describe the instantaneous state of each atom. Furthermore we have introduced the single-atom emission rate $\Gamma_c=g^2/\kappa$ into the cavity mode and collective operators for the atomic dipoles
	\begin{align}
		\hat{J}^{\pm}=\sum_j\eta({\bf x}_j)\hat{\sigma}_j^{\pm}.
	\end{align}
	The summation runs over all atoms in the beam. The effect of the shot noise that is present in this system is apparent in the terms given by ${\hat{\mathcal{S}}_j^{-}=\eta(\hat{\bf x}_j)\hat{\sigma}^z_j\hat{\mathcal{F}}^{-}}$ and ${\hat{\mathcal{S}}_j^{z}=-2\eta(\hat{\bf x}_j)(\hat{\mathcal{F}}^{+}\hat{\sigma}^{-}_j+\hat{\sigma}^{+}_j\hat{\mathcal{F}}^{-})}$. The term $\hat{\mathcal{F}}^{-}$ is effectively delta-correlated on the slow timescale associated with the dynamics of the atomic degrees of freedom. This property is represented by the set of correlations that can be written as $\langle \hat{\mathcal{F}}^{-}(t)\hat{\mathcal{F}}^{-}(t')\rangle_c=0=\langle\hat{\mathcal{F}}^{+}(t)\hat{\mathcal{F}}^{-}(t')\rangle_c$ and $\langle \hat{\mathcal{F}}^{-}(t)\hat{\mathcal{F}}^{+}(t')\rangle_c=\Gamma_c\delta(t-t')$, $\hat{\mathcal{F}}^{+}=(\hat{\mathcal{F}}^{-})^{\dag}$. The expectation value $\langle\,\ldots\,\rangle_c$ is taken over the cavity degrees of freedom and the vacuum electromagnetic modes external to the cavity. In our treatment we have neglected spontaneous emission and other dephasing mechanisms, since we assume that $\tau$ is much shorter than any single-atom decoherence time. Furthermore we assume that the atomic motion is ballistic, which is an approximation that is valid when optomechanical forces can be ignored. This requires that ${\bf F}\tau/m\ll{\bf v}$, where we can estimate the optomechanical force ${\bf F}\approx\hbar N\Gamma_c\nabla_{\bf x}\eta({\bf x})$ that is acting on an individual atom during its transit. Here, $m$ is the mass of the atom and $\nabla_{\bf x}=(\partial_x,\partial_z)$ is the gradient operator.
	
	\subsection{Semiclassical description of the atomic degrees of freedom}
	We are interested in the $N\gg1$ limit where many atoms couple to the cavity mode at the same time. Because of the exponentially large Hilbert space dimension an exact solution of the quantum mechanical Heisenberg-Langevin equations is intractable. Therefore we make a semiclassical approximation where we replace the quantum operators by $c$-numbers and add fluctuating noise terms that account for the true quantum noise. This can be done by writing down the Heisenberg-Langevin equations for the Hermitian dipole components
	$\hat{\sigma}^x_j=\hat{\sigma}^{-}_j+\hat{\sigma}^{+}_j$, $\hat{\sigma}^y_j=i(\hat{\sigma}^{-}_j-\hat{\sigma}^{+}_j)$, and $\hat{\sigma}^z_j$, and replacing them by their corresponding $c$-number variables $s_j^{x}$, $s_j^{y}$, and $s_j^{z}$. This results in the following stochastic differential equations that completely characterize our model~\cite{footnote1}
	\begin{align}
		\frac{ds_j^{x}}{dt}=&\frac{\Gamma_c}{2}\eta({\bf x}_j)s_j^{z}J_{x}+\mathcal{S}^{x}_j,\label{sxHL}\\
		\frac{ds_j^{y}}{dt}=&\frac{\Gamma_c}{2}\eta({\bf x}_j)s_j^{z}J_{y}+\mathcal{S}^{y}_j,\label{syHL}\\
		\frac{ds_j^{z}}{dt}=&-\frac{\Gamma_c}{2}\eta({\bf x}_j)\left(J_{x}s_j^{x}+J_{y}s_j^{y}\right)+\mathcal{S}^{z}_j,\label{szHL}\\
		\frac{d{\bf x}_j}{dt}=&{\bf v}_j\label{x}.
	\end{align}
	The expressions
	\begin{align}
		J_a=\sum_j\eta({\bf x}_j)s_j^a\label{Ja1},
	\end{align}
	with $a\in\{x,y\}$ define the $x$ and $y$ components of the collective dipole. 
	In this semiclassical description, the cavity vacuum noise is represented by the terms ${\mathcal{S}^{a}_j=\eta({\bf x}_j)s^z_j\mathcal{F}_{a}}$ and ${\mathcal{S}^{z}_j=-\eta({\bf x}_j)(s^x_j\mathcal{F}_{x}+s^y_j\mathcal{F}_{y})},$ where $\mathcal{F}_{x}$ and $\mathcal{F}_{y}$ have zero mean and are defined by the correlation matrix elements ${\langle \mathcal{F}_{a}(t)\mathcal{F}_{b}(t')\rangle=\Gamma_c\delta_{ab}\delta(t-t')}$ with $a,b\in\{x,y\}$ and $\delta_{ij}$ the Kronecker delta. In our approach, these noise terms have been derived using the symmetric ordering of the operators, where we identify the symmetric ordered moment 
	$\langle \hat{\sigma}_i^{a} \hat{\sigma}_j^{b}+\hat{\sigma}_j^{b} \hat{\sigma}_i^{a}\rangle/2$ as the second moment $\langle s_i^{a}s_j^{b}\rangle$ of the classical $c$-number variables. Besides the fluctuations arising from the cavity vacuum noise (i.e., $\mathcal{F}_{x}$ and $\mathcal{F}_{y}$), there are additional noise source terms that arise from the effective pumping that is introduced by atoms sporadically entering and leaving the cavity mode. For atom $j$ that enters in $|e\rangle$ with $s_j^z = 1$, the uncertainty in $s_j^x$ and $s_j^y$ needs to be maximal (see Ref.~\cite{Schachenmayer:2015}). This is modeled by randomly and independently initializing $s_j^x=\pm1$ and $s_j^y=\pm 1$. With this we fulfill the boundary conditions for the preexcited dipoles as they enter the cavity, i.e., $\langle \hat{\sigma}_j^x\hat{\sigma}_i^x\rangle=\langle s_j^xs_i^x\rangle=\delta_{ij}$, $\langle \hat{\sigma}_j^y\hat{\sigma}_i^y\rangle=\langle s_j^ys_i^y\rangle=\delta_{ij}$, and $\langle\hat{\sigma}_j^x\hat{\sigma}_i^y+\hat{\sigma}_i^y\hat{\sigma}_j^x\rangle/2=\langle s_j^xs_i^y\rangle=0$.
	
	While the microscopic description of Eqs.~\eqref{sxHL}--\eqref{x} is used for the numerical analysis of the setup, we can also derive a macroscopic description that allows for (semi)analytical results. To obtain this macroscopic description of the atomic beam we examine the dynamics of the densities ${s_a({\bf x},t)=\sum_js^{a}_j\delta({\bf x}-{\bf x}_j)}$ with ${a\in\{x,y,z\}}$. Using Eqs.~\eqref{sxHL}--\eqref{x} we obtain Klimontovich-like stochastic  equations~\cite{Campa:2009} for the densities
	\begin{align}
		\frac{\partial s_x}{\partial t}+{\bf v}\cdot\nabla_{\bf x}s_{x}=&\frac{\Gamma_c}{2}\eta({\bf x})J_xs_z+\mathcal{S}_x\label{sx}\,,\\
		\frac{\partial s_y}{\partial t}+{\bf v}\cdot\nabla_{\bf x}s_{y}=&\frac{\Gamma_c}{2}\eta({\bf x})J_ys_z+\mathcal{S}_y\,,\label{sy}\\
		\frac{\partial s_z}{\partial t}+{\bf v}\cdot\nabla_{\bf x}s_{z}=&-\frac{\Gamma_c}{2}\eta({\bf x})\left(J_xs_x+J_ys_y\right)+\mathcal{S}_z\,.\label{sz}
	\end{align}
	The left-hand sides of Eqs.~\eqref{sx}--\eqref{sz} describe the free flight of the atoms. The first term on the right-hand side of each equation characterizes the collective decay mediated by the cavity field. In this density notation the $x$ and $y$ components for the collective dipole defined in Eq.~\eqref{Ja1} can be expressed as
	\begin{align}
		J_a= \int d{\bf x}\,\eta({\bf x})s_a({\bf x},t),
	\end{align} where we have used $\int d{\bf x}\,f({\bf x})=\int_{-\infty}^{\infty} dx \int_{-\infty}^{\infty} dz\,f(x,z)$ and $a\in\{x,y\}$. 
	The $\mathcal{S}_a$ terms in Eqs.~\eqref{sx}--\eqref{sz} are stochastic variables that are described by $\mathcal{S}_a({\bf x},t)=\eta({\bf x})\mathcal{F}_as_z$  and $\mathcal{S}_z({\bf x},t)=-\eta({\bf x})\left(\mathcal{F}_xs_x+\mathcal{F}_ys_y\right)$.
	
	While the derivation so far is quite general, our analytical and numerical analyses focus on a simplified cavity mode function with a rectangular profile that is given explicitly by the form
	\begin{align}
		\eta({\bf x})=\cos(k_cz)\left[\Theta(x+w)-\Theta(x-w)\right].
	\end{align}
	Here, $\Theta(x)$ is the Heaviside step function, $w$ is a width parameter that effectively corresponds to the cavity beam waist, and $k_c=2\pi/\lambda$ is the wavenumber with $\lambda$ the optical wavelength. The transit time is directly related to the cavity beam waist and the velocity vertical to the cavity axis, i.e., $\tau=2w/v_x$. The prescribed condition that new atoms are introduced in state $|e\rangle$ leads to a boundary condition ${s_z(x=-w,z,t)=N/(2w\lambda)}.$ This is derived assuming that the diameter of the atomic beam is much larger than the wavelength $\lambda$. In this case we can use $\lambda$-periodic boundary conditions in the $z$ direction and restrict the $z$ values to the interval $[0,\lambda)$. In order to describe the quantum fluctuations of the introduced dipoles it is necessary to establish the correct magnitudes of the second moments~\cite{Schachenmayer:2015}. This results in initializing the $s_x$ and $s_y$ components with the aid of a simulated noise process that is defined by the following properties: ${s_a(x=-w,z,t)~=W_a(z,t)}$, with ${\langle W_a(z,t)\rangle=0}$ and ${\langle W_a(z,t)W_b(z',t')\rangle=N/(2w\lambda)\delta_{ab}\delta(z-z')\delta(t-t')/v_x}$, $a,b\in\{x,y\}$. 
	
	In the following section we will use this density description to study the onset of superradiance.
	\section{Onset of superradiance}\label{sec:OSR}
	We first solve Eqs.~\eqref{sx}--\eqref{sz} within the scope of a mean-field approximation. That is, we assume $s_a\approx\langle s_a\rangle$, ${a\in\{x,y,z\}}$, and calculate the expectation values of the individual dipole components. For clarity, here the expectation value $\langle\,\ldots\,\rangle$ denotes an average over different initializations and noises. By replacing the fluctuating variables $s_a$ and $J_a$ by their expectation values, we obtain the mean-field description
	\begin{align}
		\frac{\partial \langle s_x\rangle}{\partial t}+{\bf v}\cdot\nabla_{\bf x}\langle s_{x}\rangle=&\frac{\Gamma_c}{2}\eta({\bf x})\langle J_x\rangle \langle s_z\rangle,\label{meanfield sx}\\
		\frac{\partial \langle s_y\rangle }{\partial t}+{\bf v}\cdot\nabla_{\bf x}\langle s_{y}\rangle =&\frac{\Gamma_c}{2}\eta({\bf x})\langle J_y\rangle \langle s_z\rangle,\label{meanfield sy}\\
		\frac{\partial \langle s_z\rangle}{\partial t}+{\bf v}\cdot\nabla_{\bf x}\langle s_{z}\rangle =&-\frac{\Gamma_c}{2}\eta({\bf x})\left[\langle J_x\rangle \langle s_x\rangle+\langle J_y\rangle \langle s_y\rangle\right].\label{meanfield sz}
	\end{align}
	Without any noise, the system will always remain in a non-superradiant configuration ${\langle s_x\rangle=0=\langle s_y\rangle}$, and consequently ${\langle J_x\rangle=0=\langle J_y\rangle}$. In this case the atoms essentially do not interact with the cavity and there is no emission of photons. Therefore, during the transit the atoms remain in their electronic excited state, i.e.,
	\begin{align}
		\langle s_{z}\rangle=&\frac{N}{2w\lambda}.
	\end{align}
	However, this mean-field solution is in general not stable with respect to perturbations by the physical noise sources. Fluctuations of the dipoles and cavity shot noise would initiate a transient avalanche emission process and lead to collective emission by the dipoles into the cavity mode. In order to find the threshold for this effect we calculate the stability of the non-superradiant solution with respect to a small fluctuation $\delta s_a=s_a-\langle s_a\rangle$, $a\in\{x,y\}$. The equations for $\delta s_a$ read
	\begin{align}
		\frac{\partial \delta s_a}{\partial t}+{\bf v}\cdot\nabla_{\bf x}\delta s_a=&\frac{N\Gamma_c}{4w\lambda}\eta({\bf x})\delta J_a.\label{deltas}
	\end{align}
	Here, we have defined $\delta J_a=\int d{\bf x}\eta({\bf x})\delta s_a$ and neglected second order terms in the fluctuations.
	Using the Laplace transformation $L[f](\nu)=\int_{0}^\infty e^{-\nu t}f(t)dt,$ we find
	\begin{align}
		L[\delta J_a]=\frac{\int d{\bf x}\,\int_{0}^\infty dt\,e^{-\nu t}\eta({\bf x}+{\bf v}t)\delta s_a({\bf x},0)}{1-\frac{N\Gamma_c}{4w\lambda}\int d{\bf x}\,\int_{0}^\infty dt\,e^{-\nu t}\eta({\bf x}+{\bf v}t)\eta({\bf x})},\label{deltaJa}
	\end{align}
	where $\delta s_a({\bf x},0)$ is the fluctuating initial condition and we have used the notation ${\int d{\bf x}f({\bf x})=\int_{-w}^{w} dx\int _{0}^{\lambda} dz f(x,z)}$ for any function $f({\bf x})=f(x,z)$. We have provided more complete details of the derivation in Appendix~\ref{App:A}. The inverse transform back into the time domain would provide the solution for $\delta J_a$. However, what we are interested in here is the stability of this solution, that is, whether $\delta J_a$ is exponentially damped or exponentially grows. This behavior can be studied directly using the dispersion relation, i.e., the denominator of Eq.~\eqref{deltaJa}, whose roots determine the exponents in the time domain. The dispersion relation reads 
	\begin{align}
		D(\nu)=&1-\frac{N\Gamma_c}{4w\lambda}\int d{\bf x}\,\int_{0}^\infty dt\,e^{-\nu t}\eta({\bf x}+{\bf v}t)\eta({\bf x}).\label{Dishom}
	\end{align}
	The long-time behavior of $\delta J_a\propto e^{\nu_0t}$ is determined by the root $\nu_0$  of $D(\nu)$ with the largest real part. If $\nu_0$ has a negative real part the non-superradiant state is stable and $\nu_0$ determines the decay rate of fluctuations. On the other hand, if $\nu_0$ has a positive real part the fluctuations will exponentially grow and thereby seed a superradiant emission from the ensemble. 
	
\begin{figure}[h!]
		\center
		\includegraphics[width=0.9\linewidth]{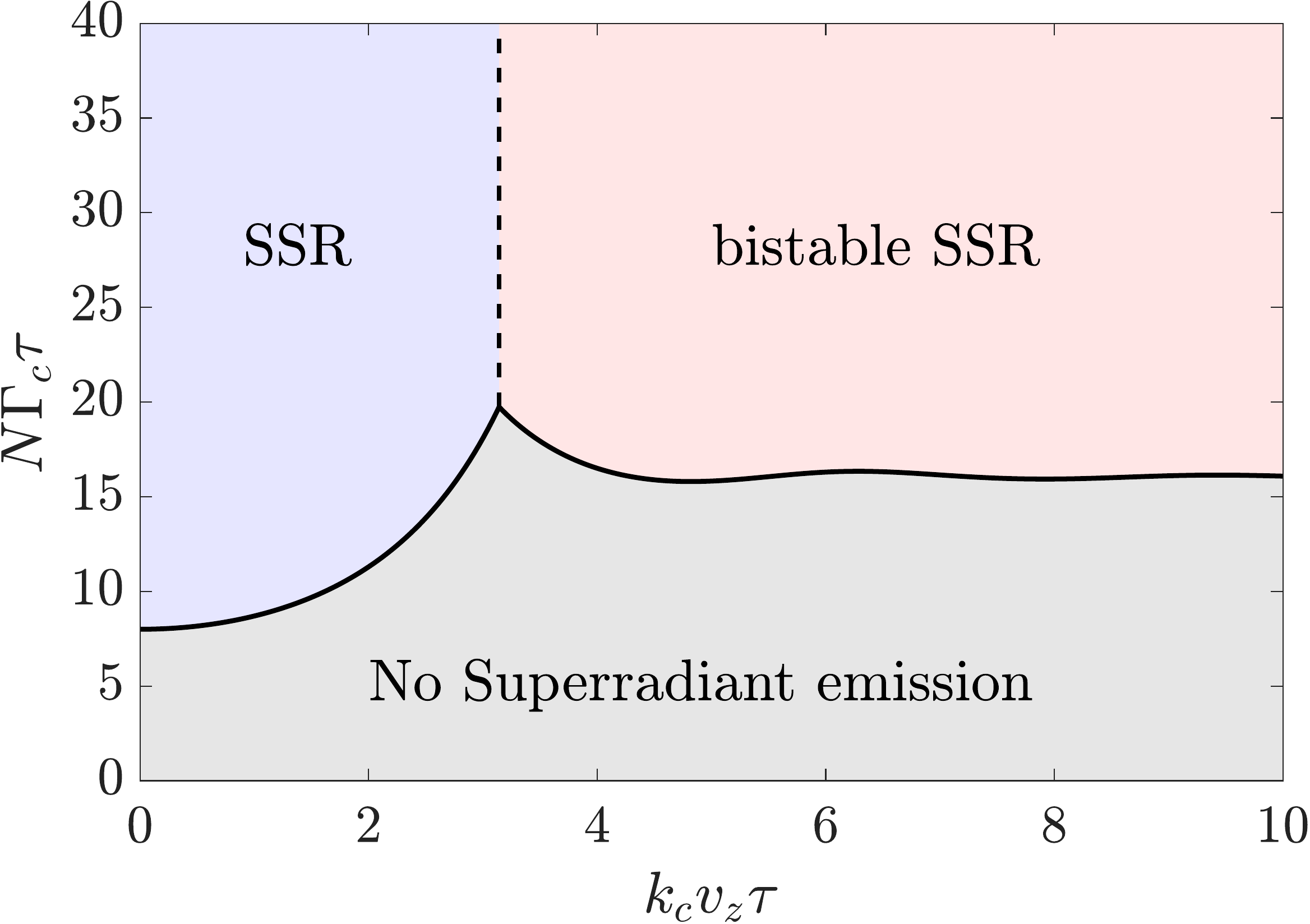}
		\caption{The resulting phase diagram describing the light emission for different values of the Doppler shift, $k_cv_z$, and the collective linewidth, $N\Gamma_c$, both in units of the inverse transit time, $1/\tau$. For small values of $N\Gamma_c\tau$ we find no superradiant emission. For sufficiently large values of $N\Gamma_c\tau$, regimes of either regular steady-state superradiance (SSR) or bistable SSR are observed depending on the magnitude of $k_cv_z\tau$. \label{Fig:2}}
	\end{figure}

	The boundary between the regime of no superradiant emission and that of superradiant emission is visible in Fig.~\ref{Fig:2} as a solid black line. This black line has been calculated by finding the roots $\nu_0$ of Eq.~\eqref{Dishom} with $\mathrm{Re}(\nu_0)=0$.
	As visible in Fig.~\ref{Fig:2}, superradiant emission emerges when the transit time broadening $~1/\tau$ is small compared to the collective linewidth $N\Gamma_c$. The exact threshold between no superradiant emission and superradiant emission depends on how many wavelengths an atom traverses during its transit. This quantity is shown as the $x$ axis in Fig.~\ref{Fig:2} that represents $k_cv_z\tau=2\pi\times (v_z\tau)/\lambda$. However, superradiance can be observed for every $v_z$ as long as $N\Gamma_c\tau>20$. 
	
	While in this section, we have been primarily concerned with the difference between superradiant and no superradiant emission, we also show in Fig.~\ref{Fig:2} two different superradiant phases. In the next section we will explain how we distinguish between these two superradiant phases and provide a detailed analysis for parameters that cross the transition boundary that separates them.
	
	\section{Superradiant phases}\label{sec:SR}
	We now focus entirely on the superradiant emission regime. In particular, we are interested in understanding the effect of $v_z$ along the cavity axis that leads to a transverse Doppler shift in the frequency of emitted photons. For a single atom, the emission of photons into the direction of motion shifts the frequency to the blue  of the atomic resonance frequency $\omega_a$, while emission in the opposite direction shifts the frequency to the red. In the following subsection we will demonstrate that this simple single-atom picture is inadequate to describe the collective system.
	\subsection{Regular SSR and bistable SSR}
\begin{figure}[h!]
		\center
		\includegraphics[width=0.85\linewidth]{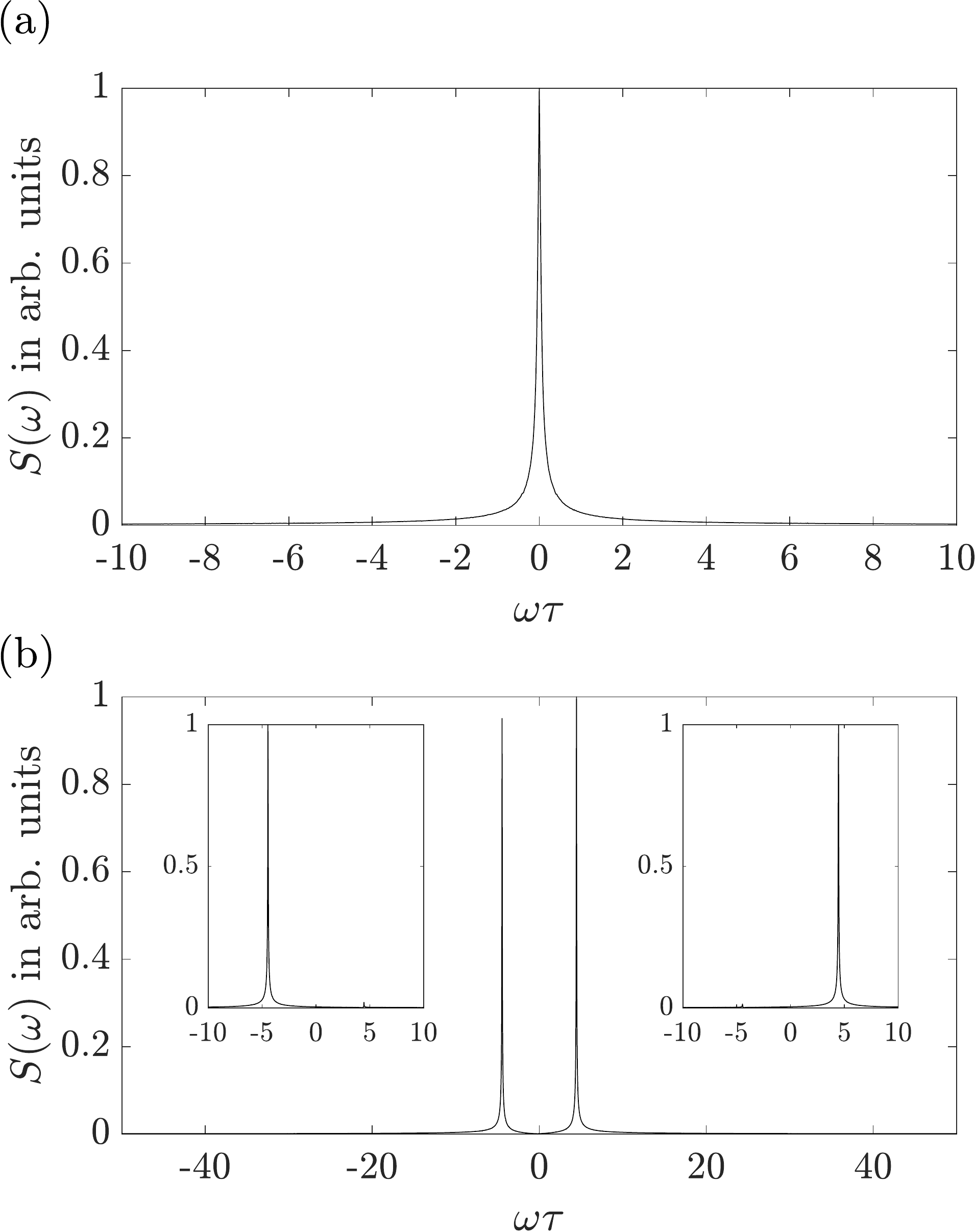}
		\caption{The spectrum $S(\omega)$, defined in Eq.~\eqref{S}, as a function of the frequency $\omega$ in units of $1/\tau$ in the SSR phase for $k_cv_z\tau=2\pi\times0.3$ (a) and in the bistable SSR phase $k_cv_z\tau=2\pi\times0.8$ (b). For the simulation we used $N\Gamma_c\tau=30$, $N=800$, and a total integration time of $t_0+T=t_{\mathrm{sim}}=2000\tau$. The spectra are calculated using 500 independent initializations and after a time $t_0=10\tau$ (after which the system is well described as being in steady state). The two insets in subplot (b) show the averaged spectrum of the trajectories that correspond to a negative frequency $\omega\tau\approx-4.46$ (238~trajectories) and positive frequency $\omega\tau\approx4.46$ (262~trajectories).
			\label{Fig:3}}
	\end{figure}

	In order to study the regimes of coherent emission, we integrate the stochastic differential equations~\eqref{sxHL}--\eqref{x} numerically for various parameters. In general, we observe that for small velocities $v_z$ the atomic beam undergoes superradiant emission that is still resonant with the bare atomic resonance frequency. This finding highlights the many-body character of the superradiant atomic beam since one might expect a Doppler-shifted frequency for the single-atom case. In order to demonstrate this behavior, we show the spectrum (see Fig.~\ref{Fig:3}a)
	\begin{align}
		S(\omega)\propto\left|\int_{0}^{T}dt\,e^{i\omega t}\langle J^*(t+t_0)J(t_0)\rangle\right|,\label{S}
	\end{align}  
	where $t_0\gg\tau$ is a sufficiently large time after which the system has evolved to a stationary state, and ${J(t)=[J_x(t)-i J_y(t)]/2}$. The time $T$ is the integration time after $t_0$ (see caption of Fig.~\ref{Fig:3}). For $k_cv_z\tau=2\pi\times0.3$, i.e., when each atom traverses 0.3 wavelengths along the cavity axis during the transit time, the spectrum shows a narrow Lorentzian peak at $\omega=0$ corresponding to continuous superradiant emission with central frequency $\omega_a$. We label this phase as SSR, due to the similarities with regular steady-state superradiance (Fig.~\ref{Fig:2}).   
	
	While this behavior remains stable at first as $v_z$ is increased, once a critical velocity is reached we observe a threshold beyond which a qualitatively different behavior emerges. As an example, we show $S(\omega)$ for $k_cv_z\tau=2\pi\times0.8$ in Fig.~\ref{Fig:3}b, corresponding to each atom traversing 0.8 wavelengths along the cavity axis. In this case, the spectrum exhibits two narrow Lorentzian peaks that are symmetrically shifted from the resonance frequency of the atoms. While the form of the spectrum suggests simultaneous emission with both frequencies, we find that the atomic beam will randomly undergo superradiant emission with either the red or the blue detuned frequency. The random choice is seeded by the first emission with probability of 0.5 for each of the two possibilities. Subsequently collective spontaneous emission events will amplify the light field with that frequency. 
	
	To further demonstrate this behavior, we illustrate in the left (right) inset of Fig.~\ref{Fig:3}b the emission spectrum corresponding to trajectories that emit with red (blue) detuned frequencies. Since we have a finite number of initializations we may observe a slight imbalance of red-detuned frequencies with respect to blue-detuned frequencies in each trial batch. This imbalance can be seen as different heights in the spectrum shown in Fig.~\ref{Fig:3}b. In the insets we see only one peak supporting our claim that superradiant emission appears for the shown parameters only on one sideband. Because of the bistable nature of the superradiant peaks, this is reminiscent of optical bistability of intensity solutions~\cite{Abraham:1982}, and consequently we refer to this phase as bistable SSR (Fig.~\ref{Fig:2}). 
	
\begin{figure}[h!]
		\center
		\includegraphics[width=0.85\linewidth]{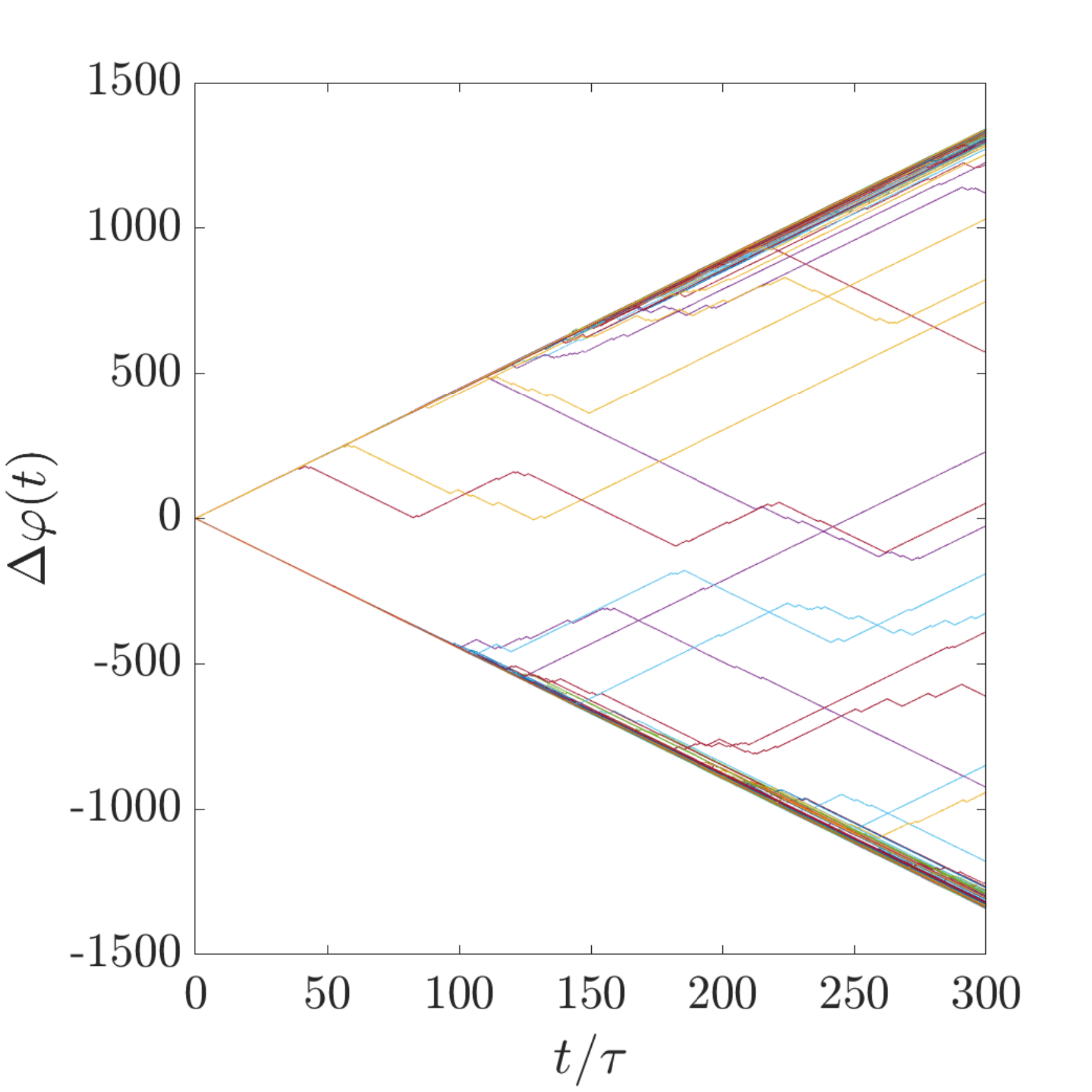}
		\caption{The phase difference $\Delta\varphi(t)$, defined in Eq.~\eqref{Deltaphi}, as a function of time in units of $\tau$ for $N=800$. The time window is defined by $t_0=10\tau$ and $t_1=1700\tau$, and the total simulation time is $t_{\mathrm{sim}}=2000\tau$. For the simulations we used $500$ trajectories and the parameters $k_cv_z\tau=2\pi\times0.8$ and $N\Gamma_c\tau=30$.\label{Fig:4}}
	\end{figure}

	This bistable behavior is best visible in the dynamics of the phase 
	\begin{align}
		\Delta \varphi(t)=\mathrm{arg}\left(\int_{t_0}^{t_1} dt_0'\frac{\left\langle J^*(t+t_0')J(t_0')\right\rangle}{t_1-t_0}\right),\label{Deltaphi}
	\end{align}
	where $\mathrm{arg}(\,\ldots\,)$ denotes the argument and $t_0$ and $t_1$ are the initial and final times of an averaging window. We show the dynamics of the phase $\Delta\varphi$ in Fig.~\ref{Fig:4} with 500 initializations and for the same parameters as in Fig.~\ref{Fig:3}b, $N\Gamma_c\tau=30$ and $k_cv_z\tau=2\pi\times0.8$.
	Most of the 500 trajectories remain on straight lines with a constant slope. This slope corresponds to the two frequencies that are visible in Fig.~\ref{Fig:3}b. However, some of the trajectories jump between the two slopes, signifying clearly the bistable nature of the frequency solutions. 
	
	In order to understand further properties of the two superradiant phases and to provide insight that is evident from an analytic treatment, we now develop a mean-field theoretic description.
	
	\subsection{Intensity and emission frequency}
	Both superradiant phases can be classified by a non-vanishing collective dipole with a constant length. However, in one phase the collective dipole oscillates with a non-vanishing frequency $\omega$ (bistable SSR) while in the other regime the phase of the collective dipole remains almost constant (regular SSR).
	
	In order to analyze this behavior we solve the mean-field equations 
	\begin{align}
		\frac{\partial \langle s\rangle}{\partial t}+{\bf v}\cdot\nabla_{\bf x}\langle s\rangle=&\frac{\Gamma_c}{2}\eta({\bf x})\langle s_z\rangle \langle J\rangle,\label{s}\\
		\frac{\partial \langle s_z\rangle}{\partial t}+{\bf v}\cdot\nabla_{\bf x}\langle s_z\rangle=&-\Gamma_c\eta({\bf x})\left[\langle J^*\rangle\langle s\rangle+\langle s^*\rangle \langle J\rangle\right],\label{z}
	\end{align}
	that are presented in the form above for the complex dipole ${s=(s_x-is_y)/2}$ with $ J=\int d{\bf x}\,s$. From Eqs.~\eqref{s}--\eqref{z} one can verify that
	\begin{align}
		\left(\frac{\partial}{\partial t}+{\bf v}\cdot\nabla_{\bf x}\right)\left[\langle s_z\rangle^2+4|\langle s\rangle |^2\right]=0.
	\end{align}
	This equation highlights that in our model the length of the Bloch vector is conserved. This is a consequence of the form of Eqs.~\eqref{s}--\eqref{z} that describe collective emission as Rabi oscillations with a self-consistent Rabi frequency~$\propto\langle J\rangle$. As a result we can use spherical coordinates to describe the dipole densities. Together with the boundary conditions, we therefore parametrize the spin variables by the following geometrical quantities
	\begin{align}
		\langle s\rangle=&\frac{N}{4w\lambda}e^{-i\phi({\bf x},t)}\sin\left(K({\bf x},t)\right),\\
		\langle s_z\rangle=&\frac{N}{2w\lambda}\cos\left(K({\bf x},t)\right),
	\end{align}
	with space and time dependent angles $\phi({\bf x},t)$ and $K({\bf x},t)$. 
	
	While this description is always valid we will now focus on the stationary properties of the atomic beam that are realized after a sufficiently long time $t$. In both regular SSR and bistable SSR, we anticipate a behavior for $\phi({\bf x},t)$ according to
	\begin{align}
		\phi({\bf x},t)=\omega t+\psi({\bf x}), \label{phi}
	\end{align} 
	where $\omega$ is the frequency of the emitted light and $\psi$ is a position dependent but time independent phase. Assuming $K$ is not explicitly time dependent, we obtain the following coupled differential equations for $\psi$ and $K$
	\begin{align}
		\omega+{\bf v}\cdot\nabla_{\bf x}\psi=&-\Gamma_c\eta({\bf x})|\langle J\rangle|\sin(\psi)\cot\left(K\right),\label{psi}\\
		{\bf v}\cdot\nabla_{\bf x}K=&\Gamma_c\eta({\bf x})|\langle J\rangle |\cos(\psi).\label{K}
	\end{align}
	These equations can be solved together with the two equations emerging from the real and imaginary parts of $\int d{\bf x} \langle s\rangle e^{i\omega t}=|\langle J\rangle|$. The solution of all four equations result in a value for the length of the collective dipole $|\langle J\rangle|$, the emission frequency $\omega$, and the functions $K({\bf x})$ and $\psi({\bf x})$.  We have derived these equations, without loss of generality, under the assumption that $\langle J(t=0)\rangle=\langle J_x(t=0)\rangle/2$ points in the $x$ direction at $t=0$. This is equivalent to the assumption $\langle J\rangle=|\langle J\rangle|e^{-i\omega t}$. The complexity in solving Eqs.~\eqref{psi}--\eqref{K} is tremendously simplified in the case where $\omega=0$ (regular SSR phase) because we directly obtain the result $\psi=0$. We report the solution of this equation for the case $\omega=0$ in Appendix~\ref{App:B}. However, for the general case we have to solve the coupled partial differential equations.
	
\begin{figure}[h!]
		\center
		\includegraphics[width=1\linewidth]{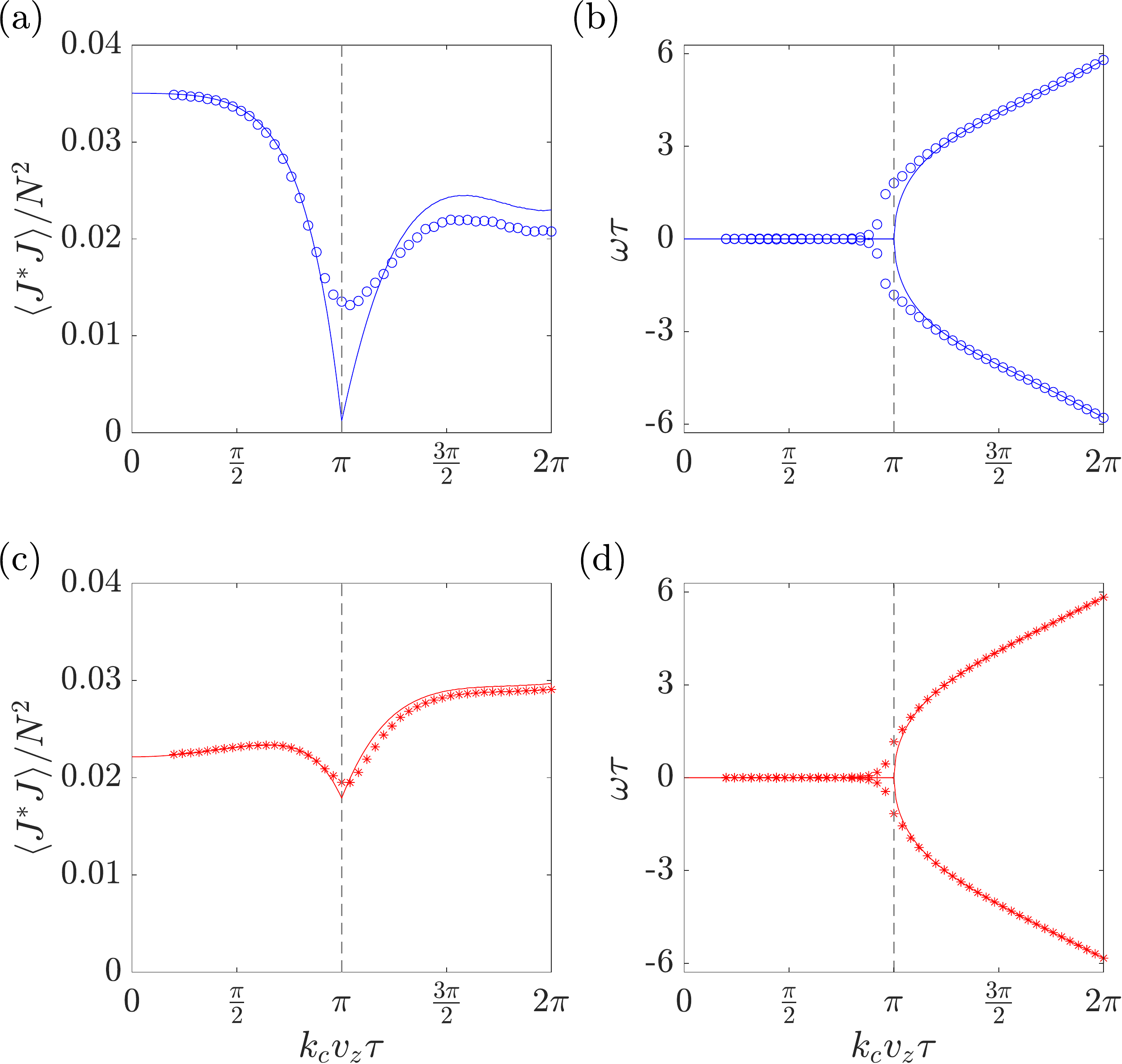}
		\caption{ The collective dipole $\langle J^*J\rangle/N^2$, subplots (a--c), and the frequency of the light $\omega$ in units of $1/\tau$, subplots (b--d), as functions of $k_cv_z\tau$. Subplots (a--b), and (c--d) show results for ${N\Gamma_c\tau=20}$ and ${N\Gamma_c\tau=30}$, respectively. The circles and stars correspond to numerical simulations of Eqs.~\eqref{sxHL}--\eqref{x}, and the solid lines represent analytical solutions for $N\to\infty$. The vertical gray dashed lines show the transition from regular SSR to bistable SSR. The numerical values of $\omega$ from the simulations in subplots (b) and (d) have been calculated by fitting $g_1(t)$ (Eq.~\eqref{g1normalized}) to ${\cos(\omega t+\phi_0)e^{-\Gamma t/2}}$ and $t_0=10\tau$. Here, $\omega$, $\Gamma$ and $\phi_0$ are fitting parameters. The simulations are performed with $N=800$, an integration time of $t_{\mathrm{sim}}=100\tau$, and $400$ initializations.\label{Fig:5}}
	\end{figure}

	We show the mean-field results for $\omega$ and $|\langle J\rangle|$ across the regular SSR to bistable SSR transition and compare them  with the results of a numerical integration of Eqs.~\eqref{sxHL}--\eqref{x}. The results are calculated for $N\Gamma_c\tau=20$ visible in Fig.~\ref{Fig:5}a--b, close to the non-superradiant regime, and for $N\Gamma_c\tau=30$ shown in Fig.~\ref{Fig:5}c--d, well inside of the superradiant regime.  	
	In Fig.~\ref{Fig:5}, we illustrate $\langle J^*J\rangle/N^2$ and the emission frequency $\omega$ as a function of $k_cv_z\tau$. The mean-field theory predicts a non-analytical behavior of both $\langle J^*J\rangle/N^2$ and $\omega$ at a threshold value of $k_cv_z\tau=\pi$. It shows a kink-like local minimum for $\langle J^*J\rangle/N^2$ and a bifurcation of $\omega$ at the threshold that is in agreement with the simulations. In general we find that the non-analyticities are smoothed out by noise and finite size effects. The rather large discrepancies between the mean-field results and the simulations in Fig.~\ref{Fig:5}a are likely due to these effects that are more pronounced close to a tri-critical point where regular SSR, bistable SSR, and the non-superradiant emission phases meet (tri-critical point is at $N\Gamma_c\tau=2\pi^2$ and $k_cv_z\tau=\pi$). For the large $k_cv_z\tau$ limit we obtain the asymptotic result $\omega\approx k_cv_z$. The behavior of $\omega$ close to the transition is reminiscent of a second order phase transition that is here observed in a highly dissipative setting where neither individual atoms nor individual photons remain in the cavity on a timescale longer than $\tau$. We remark that in both superradiant phases we have broken a $U(1)$ symmetry resulting in a well defined value for the phase of $J$ and corresponding physically to the generation of near-monochromatic light. In the bistable SSR phase we also have a broken time-translation symmetry, which is evident in Eq.~\eqref{phi} for $\omega\neq0$. 
	
	As we have pointed out in the previous subsection the system can jump between the two bistable frequencies~$\pm\omega$. We now analyze the statistical properties of this effect in more detail using the result of the numerical integration of Eqs.~\eqref{psi}--\eqref{K}).
	\subsection{Mode hopping probability} 
	In order to quantitatively analyze the statistical properties of the mode hopping, we calculate the probability for the occurrence of a jump from the negative to the positive frequency. In order to do this, we begin by evaluating $\Delta\varphi(t)$ according to Eq.~\eqref{Deltaphi}. Then, we divide the time interval $[0,t_{\mathrm{max}}]$ of every trajectory of $\Delta\varphi(t)$ into $M$ equal interval time bins $[(m-1)\Delta t,m\Delta t]$ with $m=1,\ldots,M$ and $\Delta t=t_{\mathrm{max}}/M$. Within each time bin we calculate an average frequency
	\begin{align}
		\omega(m)=\frac{1}{\Delta t}\int_{(m-1)\Delta t}^{m\Delta t}dt'\,\frac{d\Delta\varphi(t')}{dt'}.
	\end{align}
	From the average frequencies, we can now accumulate statistics on the number of frequency jumps that occur by evaluating whether $\omega(m)\omega(m+1)<0$ for ${m=1,\ldots,M-1}$. By counting the total number of jumps from all trajectories, $\mathcal{N}_{\mathrm{jump}}$, and dividing by the maximum number of jumps possible, ${\mathcal{N}_{\mathrm{total}}=(M-1)\times\mathcal{T}}$, where $\mathcal{T}$ is the number of trajectories, we get
	\begin{align}
		P_{\mathrm{jump}}=\frac{\mathcal{N}_{\mathrm{jump}}}{\mathcal{N}_{\mathrm{total}}}\label{Jumpprobab}
	\end{align}
	for the probability of a mode hop.
	\begin{figure}[h!]
		\center
		\includegraphics[width=0.75\linewidth]{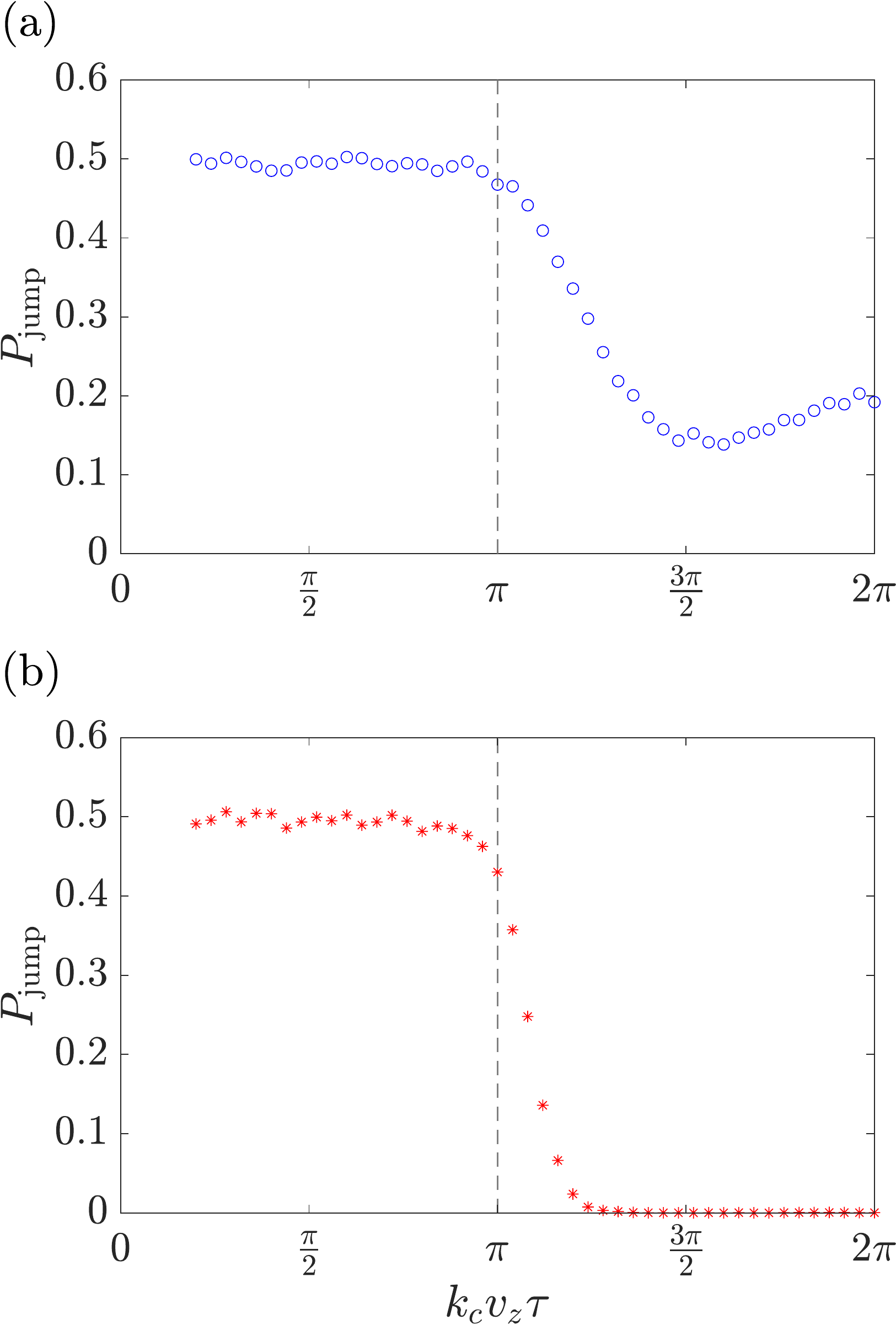}
		\caption{The jump probability $P_{\mathrm{jump}}$, defined in Eq.~\eqref{Jumpprobab}, for different values of $k_cv_z\tau$ for $N\Gamma_c\tau=20$ (a) and $N\Gamma_c\tau=30$ (b). For the simulations we used $t_{\mathrm{sim}}=100\tau$, $N=800$, and $\mathcal{T}=400$ and started the analysis after $t_{0}=10\tau$, after which, to good approximation, the system had reached the stationary state. The value of $\Delta\varphi(t)$ for each trajectory is calculated according to Eq.~\eqref{Deltaphi} without the time average for $t_1\to t_0$.  According to the definitions given in the text prior to Eq.~\eqref{Jumpprobab}, we have used $t_{\mathrm{max}}=90\tau$ that we split into $M=20$ bins. The gray dashed vertical line shows the threshold between the regular SSR and the bistable SSR phases, i.e., $k_cv_z\tau=\pi$. \label{Fig:6}}
	\end{figure}
	
	The jump probability is shown in Fig.~\ref{Fig:6}a--b for various values of $k_cv_z\tau$ across the phase transition from regular SSR to bistable SSR and for $N\Gamma_c\tau=20$ (Fig.~\ref{Fig:6}a) and $N\Gamma_c\tau=30$ (Fig.~\ref{Fig:6}b), respectively. The simulations are the same as those shown in Fig.~\ref{Fig:5}. We see that $P_{\mathrm{jump}}$  is close to $P_{\mathrm{jump}}\approx0.5$ for both values of $N\Gamma_c\tau$ well inside the regular SSR phase. This can be explained by the fact that $\Delta\varphi$ diffuses. In this case, after every time bin, the total phase gains with probability $0.5$ a positive or negative increment. Beyond the transition point, $k_cv_z\tau=\pi$, we observe a decrease of this jump probability in both cases. For $N\Gamma_c\tau=30$ (Fig.~\ref{Fig:6}b), we observe that the jump probability drops to a value very close to $P_{\mathrm{jump}}\approx0$. This emphasizes that the switch between a negative and a positive frequency becomes very improbable. While we also see a decrease of the jump probability for $N\Gamma_c\tau=20$ (Fig.~\ref{Fig:6}a), after the transition point, a jump is still much more likely than for $N\Gamma_c\tau=30$. Moreover, we observe that the jump probability shows a local minimum very close to the local maximum of the amplitude of the collective dipole (see Fig.~\ref{Fig:5}a). Therefore we propose that the reason for this effect is the more pronounced contribution of noise with respect to the mean value of the collective dipole. For the same reason we expect that the jump probability will decrease in the bistable SSR phase for larger atom number $N$ since the ratio of noise to the mean value of the collective dipole is further reduced. 
	
	While deep in the regular SSR phase we have observed a diffusive behavior of the phase $\Delta\varphi$, we have also seen a ballistic behavior inside of the bistable SSR phase (see Eq.~\eqref{phi} and Fig.~\ref{Fig:5}d). This dynamical phase transition is highlighted in the linewidth of the collectively emitted light as we show now. 
	
	\subsection{The Linewidth\label{subsec: linewidth}}
	Well inside the regular SSR phase we may assume that the system has a macroscopic collective dipole with some arbitrary phase~$\varphi$ in the $x$-$y$ plane. In that case we can rotate into a frame such that $J_{\parallel}\sim N$ and $J_{\perp}\sim\sqrt{N}$, where $\parallel$ and $\perp$ denote the new $x$ and $y$ axes. The direction corresponding to $J_{\parallel}$ is the direction of the collective dipole while the perpendicular direction $J_{\perp}$ is solely dominated by fluctuations. The dynamics of the dipole component in the perpendicular direction can be derived from Eqs.~\eqref{sx}--\eqref{sz} as
	\begin{align}
		\frac{\partial s_{\perp}}{\partial t}+{\bf v}\cdot\nabla_{\bf x}s_{\perp}\approx&\frac{\Gamma_c}{2}\eta({\bf x})J_{\perp}s_{z,\mathrm{st}}+\mathcal{S}_{\perp},\label{svert}
	\end{align}
	where we have dropped second order terms in the fluctuations and noise and are therefore able to substitute the mean-field solution for $s_{z}$ that reads
	\begin{align}
		s_{z,\mathrm{st}}=\frac{N}{2w\lambda}\cos(K({\bf x})).
	\end{align}
	Here, $K$ is the solution of Eq.~\eqref{K} for $\omega=0=\psi$ in the SSR phase. Equation~\eqref{svert} includes cavity noise described by the quantity $\mathcal{S}_{\perp}({\bf x},t)=\eta({\bf x})\mathcal{F}_\perp s_z$  with ${\langle \mathcal{F}_\perp(t)\rangle=0}$ and ${\langle \mathcal{F}_\perp(t)\mathcal{F}_\perp(t')\rangle=\Gamma_c\delta(t-t')}$. Besides the cavity noise, it also includes the noisy boundary condition that arises from the introduction of new atoms ${s_\perp(x=-w,z,t)~=W_\perp(z,t)}$, with ${\langle W_\perp(z,t)\rangle=0}$ and ${\langle W_\perp(z,t)W_\perp(z',t')\rangle=N/(2w\lambda)\delta(z-z')\delta(t-t')/v_x}.$
	
	We can integrate Eq.~\eqref{svert} to obtain an analytical result for $J_{\perp}$ (see Appendix~\ref{App:C}). Using  ${d\varphi/dt\approx J_{\parallel,\mathrm{ st}}^{-1}dJ_{\perp}/dt}$,  where $J_{\parallel,\mathrm{st}}$ is the length of the collective dipole, we can then derive an expression for the phase $\varphi(t)$. Arguing that the origin of a finite linewidth in the regular SSR phase is phase diffusion, we can calculate the linewidth using
	\begin{align}
		\Gamma =\lim_{t\to\infty}\frac{\left\langle \Delta \varphi(t)^2\right\rangle}{t},\label{Gamma}
	\end{align}
	where we defined $\Delta\varphi(t)=\varphi(t)-\varphi(0).$
	
\begin{figure}[h!]
		\center
		\includegraphics[width=0.75\linewidth]{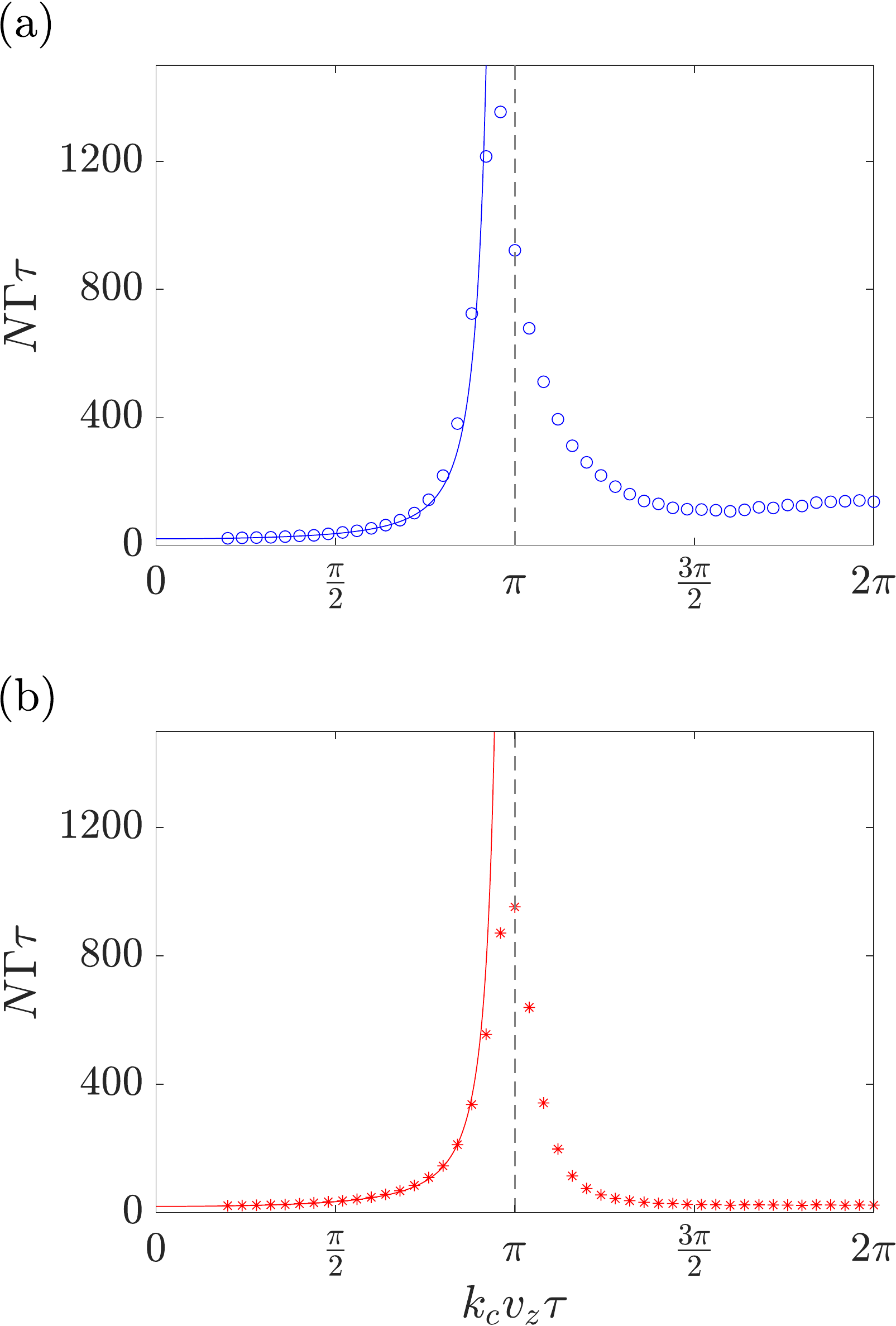}
		\caption{The linewidth $\Gamma$ in units $1/(N\tau)$ as functions of $k_cv_z\tau$ for ${N\Gamma_c\tau=20}$ (a) and ${N\Gamma_c\tau=30}$ (b). The circles and stars correspond to numerical simulations, and the solid lines represent the result of Eq.~\eqref{Gamma} for $N\to\infty$. The vertical gray dashed lines show the transition from regular SSR to bistable SSR. The values of $\Gamma$ have been calculated by fitting $g_1$ (Eq.~\eqref{g1normalized}) with ${\cos(\omega t+\phi_0)e^{-\Gamma t/2}}$ and $t_0=10\tau$. The simulations are performed with $N=800$, $400$ trajectories, and an integration time of $t_{\mathrm{sim}}=100\tau$. \label{Fig:7}}
	\end{figure}

	To show that this description is valid we have integrated numerically Eqs.~\eqref{sxHL}--\eqref{x}, calculated the real part of the normalized $g_1$ function
	\begin{align}
		g_1(t)=\frac{\mathrm{Re}\left(\left\langle J^*(t+t_0)J(t_0)\right\rangle\right)}{\langle |J(t_0)|^2\rangle},\label{g1normalized}
	\end{align}	
	and fitted ${\cos(\omega t+\phi_0)e^{-\Gamma t/2}}$ where $\omega$, $\Gamma$, and $\phi_0$ are fitting parameters. In this fit $\omega$ is the emission frequency reported in Fig.~\ref{Fig:5} and $\Gamma$ is the linewidth, visible as circles and stars in Fig.~\ref{Fig:7} for $N\Gamma_c\tau=20$~(a) and $N\Gamma_c\tau=30$~(b). The solid lines in Fig.~\ref{Fig:7} are the calculated linewidth from Eq.~\eqref{Gamma}. These curves are in good agreement with the simulations well inside of the regular SSR phase, but predict a diverging linewidth at the critical point. 
	
	The origin of this divergence in the analytical result is the break-down of the phase diffusion argument. As we show in Appendix~\ref{App:C} this divergence occurs at $k_cv_z\tau=\pi$ that we identify as the threshold between the regular SSR and bistable SSR phases. This phase boundary is shown in
	Fig.~\ref{Fig:2}, Fig.~\ref{Fig:5},
	Fig.~\ref{Fig:6}, and
	Fig.~\ref{Fig:7} as the vertical dashed lines. At this critical point, we expect that also the numerical result of the linewidth when expressed in units of $1/(N\tau)$ diverges in the large $N$ limit.
	
\begin{figure}[h!]
		\center
		\includegraphics[width=0.85\linewidth]{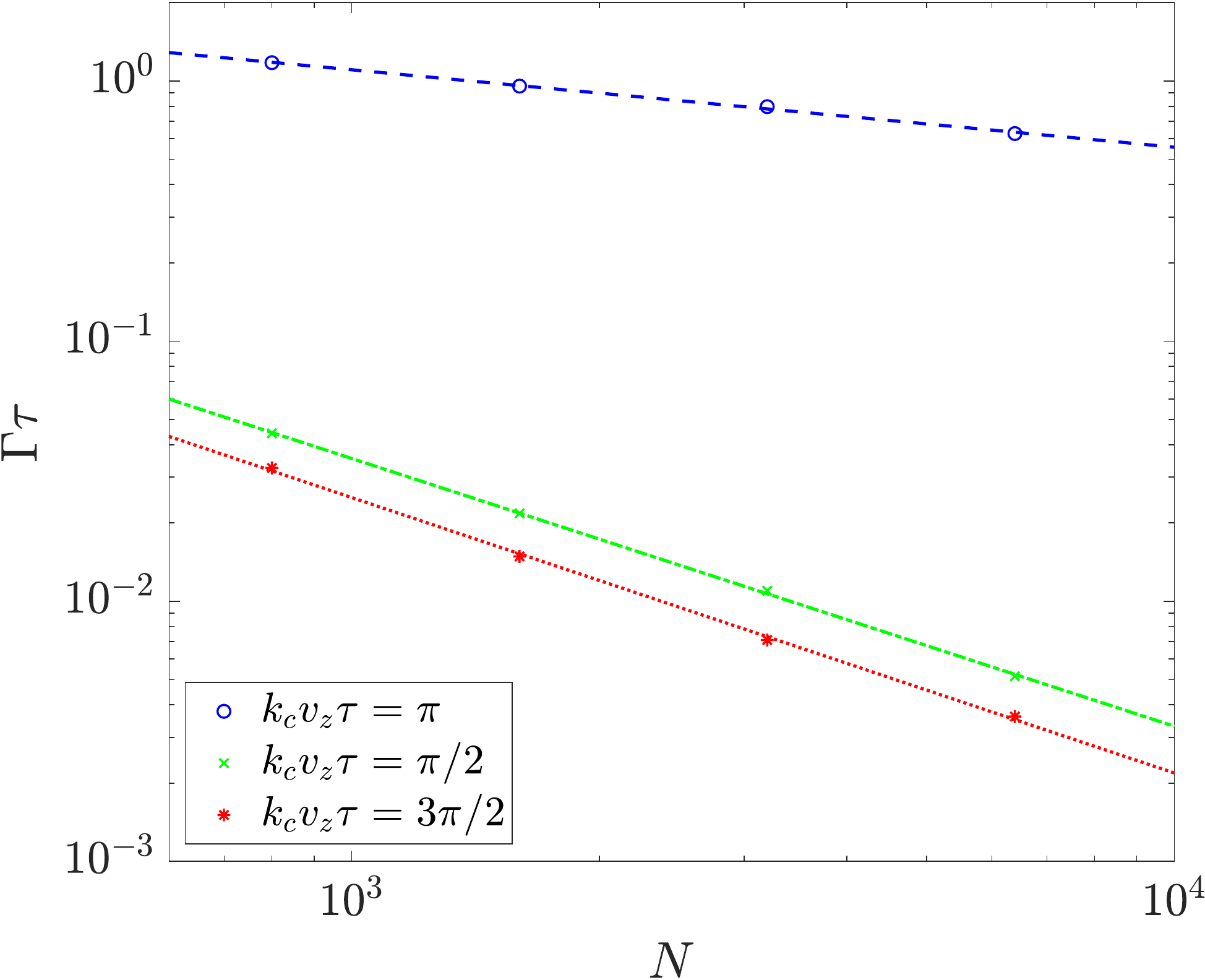}
		\caption{The linewidth $\Gamma$ in units of $1/\tau$ as a function of the intracavity atom number $N$ for $N\Gamma_c\tau=30$. The blue circles, green crosses, and red stars correspond to different values of $k_cv_z\tau$ (see legend) at the threshold, in the SSR phase, and in the bistable SSR phase. The blue dashed, green dashed-dotted, and red dotted lines are linear fits according to $\Gamma\tau\propto N^{\alpha}$ with $\alpha=-0.30$, $\alpha=-1.03$, and $\alpha=-1.06$, respectively. For every $N$ we average over $4.8\times10^5/N$ trajectories with a simulation time $t_{\mathrm{sim}}=100\tau$. Every point is calculated using the fit as described in the caption of Fig.~\ref{Fig:7}. \label{Fig:8}}
	\end{figure}

	In order to support this claim, we plot $\Gamma$ in units of $1/\tau$ for different values of $N$ in a log-log plot to illustrate the scaling of $\Gamma$ with the number of atoms, $\Gamma\tau\propto N^{\alpha}$ (Fig.~\ref{Fig:8}).
	We show the scaling well inside the regular SSR phase for $k_cv_z\tau=\pi/2$ (green crosses), well inside the bistable SSR phase for $k_cv_z\tau=3\pi/2$ (red stars), and at the theoretically predicted threshold  $k_cv_z\tau=\pi$ (blue circles).
	The values of the exponent $\alpha$ governing the scaling relation $\Gamma\tau \propto N^\alpha$ in the three regimes are extracted using a linear fit and are reported in the caption of Fig.~\ref{Fig:8}. For parameters well inside of the regular SSR or bistable SSR phases we obtain an exponent $\alpha\approx-1$. This implies that for given values of $k_cv_z\tau$ and $N\Gamma_c\tau$, $\Gamma$ in units of $1/(N\tau)$ is a constant $\propto\Gamma_c$. This claim is consistent with our theoretical description and also shows that the collective dipole remains stable on timescales that exceed the transit time $\tau$ by orders of magnitude.
	
	At the critical point $k_cv_z\tau=\pi$ the phase diffusion argument anticipates a diverging linewidth. Our numerical simulations here show that there exists a critical scaling with an exponent $\alpha\approx -0.3$. Therefore even at the critical point, $\Gamma$ beats the Fourier limit set by $1/\tau$. In units of $1/(N\tau)$ the linewidth scales as $N^{1+\alpha} \approx N^{0.7} \to \infty$, supporting our theoretical prediction of a diverging linewidth using the phase diffusion model. This divergence is reminiscent of the quantum critical region~\cite{Sachdev:2011} that occurs at finite temperature in an equilibrium quantum phase transition where scaling laws provide the potential for extreme sensitivity to model parameters.
	
	Our analytical theory can also give some insight to the relaxation dynamics at the threshold. Here, we find that 
	\begin{align}
		\langle \Delta\varphi(t)^2\rangle\propto \frac{t^3}{N},\label{t^3}
	\end{align}
	in comparison to $t/N$ inside of the SSR phase. For further details we refer to Appendix~\ref{App:D}. The superdiffusive behavior at the threshold would result in a relaxation timescale~$\sim N^{1/3}$. This result is comparable with the timescale~$\sim N^{0.3}$ that is given by the inverse of the linewidth at the threshold.  
	
	\section{Discussion and Conclusion}\label{sec:Concl}
	
	A bifurcation in the emission spectrum and a critical scaling of the linewidth has also been reported for a synchronization transition of two atomic ensembles coupled to a lossy cavity~\cite{Xu:2014,Weiner:2017}. Although the observed features may appear to be remarkably similar, we want to emphasize that the dynamical phase transition discussed here is quite different. In our model, the emission in the regular SSR and bistable SSR phases always appear with a monochromatic but possibly bistable frequency. On the other hand, the unsynchronized phase in Refs.~\cite{Xu:2014,Weiner:2017} shows a beating of two frequencies that results from simultaneous output. Moreover, the synchronization transition in Refs.~\cite{Xu:2014,Weiner:2017} appears if the collective linewidth becomes comparable to the frequency splitting of the two ensembles. Here, however, the transition between regular SSR and bistable SSR occurs if the atoms travel exactly half a wavelength during $\tau$, i.e. $k_cv_z\tau=\pi$, independent of $N\Gamma_c$. Therefore the transition from regular SSR to bistable SSR results from the dipole accumulating a phase when it travels through the cavity mode function.
	
	We emphasize that the regular SSR and bistable SSR phases rely on the continuous driving and dissipation of quantum matter, here realized by a beam of preexcited atoms. We provide the tools to analyze such systems and believe that this work will be useful as one of the first stepping stones towards future investigations of collective effects in atomic beams. For the experimental realization of such systems one requires a continuous and dense beam of atoms with a narrow transition that couples to a single cavity mode. The transition between the superradiant phases occurs when $N\Gamma_c\tau>20$ that is achievable by state-of-the-art cavity setups~\cite{Norcia:2016:1, Norcia:2016:2, Laske:2019, Gothe:2019, Schaffer:2020} combined with high phase-space density atomic beams~\cite{Chen:2019}. 
	
	Future work could investigate the regular SSR and bistable SSR phases in presence of more than just a single velocity. This includes more sophisticated models where for instance the velocity distribution is broadened. Moreover, we expect that the system is very sensitive to perturbations at the boundary between the regular SSR and bistable SSR phases. Therefore it will be interesting to investigate the potential of this system, in particular in vicinity of the critical region, for metrological applications.
	\acknowledgments
	This work was supported by the NSF PFC Grant No. PHY 1734006 and the DARPA and ARO Grant No. W911NF-16-1-0576. 
	
	\appendix
	\section{Derivation of the Dispersion function}\label{App:A}
	In this section we will describe how we can derive Eq.~\eqref{deltaJa} from Eq.~\eqref{deltas}. For this we define the operator
	\begin{align}
		\mathcal{L}_0f({\bf x)}=-{\bf v}\cdot\nabla_{\bf x}f({\bf x}), \label {L0}
	\end{align}
	with a function $f({\bf x})=f(x,z)$. With this definition, we  use the Laplace transformation on Eq.~\eqref{deltaJa} and obtain
	\begin{align}
		\left[\nu -\mathcal{L}_0\right]L[\delta s_a]-\delta s_a({\bf x},0)=\frac{N\Gamma_c}{4w\lambda}\eta({\bf x})L[\delta J_a].
	\end{align}
	We  solve this for $L[\delta s_a]$ where we obtain
	\begin{align}
		L[\delta s_a]=\left[\nu -\mathcal{L}_0\right]^{-1}\delta s_a({\bf x},0)+\frac{N\Gamma_c}{4w\lambda}L[\delta J_a]\left[\nu -\mathcal{L}_0\right]^{-1}\eta,
	\end{align}
	where we have relied on the fact that $L[\delta J_a]$ does not depend on ${\bf x}$. We can now multiply this equation by $\eta$ and integrate over ${\bf x}$ to obtain
	\begin{align}
		L[\delta J_a]=J_1+J_2,
	\end{align}
	with
	\begin{align}
		J_1=&\int d{\bf x}\eta({\bf x})\left[\nu -\mathcal{L}_0\right]^{-1}\delta s_a({\bf x},0),\\
		J_2=&\int d{\bf x}\eta({\bf x})\frac{N\Gamma_c}{4w\lambda}L[\delta J_a]\left[\nu -\mathcal{L}_0\right]^{-1}\eta.
	\end{align}
	We solve the equation for $L[\delta J_a]$ and the final result reads
	\begin{align}
		L[\delta J_a]=\frac{J_1}{1-\frac{N\Gamma_c}{4w\lambda}\int d{\bf x}\eta({\bf x})\left[\nu -\mathcal{L}_0\right]^{-1}\eta({\bf x})}.
	\end{align}
	Using now the relations
	\begin{align}
		\left[\nu -\mathcal{L}_0\right]^{-1}=&\int_{0}^\infty dt e^{-\nu t}e^{\mathcal{L}_0t},\label{trick1}\\
		e^{\mathcal{L}_0t}f({\bf x})=&f({\bf x}-{\bf v}t),\label{trick2}
	\end{align}
	and after a substitution ${\bf x}\mapsto{\bf x}+{\bf v}t$ we obtain the result in Eq.~\eqref{deltaJa}.
	
	\section{The dipole density in the regular SSR phase}\label{App:B}
	The purpose of this section is to present the analytical result of the coupled Eqs.~\eqref{psi}--\eqref{K} in the regular SSR phase where $\omega=0$. In this case Eq.~\eqref{psi} can be directly solved using $\psi=0$ and this results in the partial differential equation
	\begin{align}
		{\bf v}\cdot\nabla_{\bf x}K=\Gamma_c\eta({\bf x})|\langle J\rangle |.\label{DGLK}
	\end{align}
	The solution of this equation is straight forward and reads
	\begin{align}
		K(x-w,z)=\frac{\Gamma_cJ_{\parallel,\mathrm{ st}}\sin\left(\frac{v_z}{2v_x}k_cx\right)\cos\left(k_c\left[z-\frac{v_z}{2v_x}x\right]\right)}{k_cv_z},\label{Kanalytic}
	\end{align}
	where we have used 
	\begin{align}
		J_{\parallel,\mathrm{st}}=&2|\langle J\rangle|\nonumber\\
		=&\int d{\bf x}\,\eta({\bf x})\frac{N}{2w\lambda}\sin(K({\bf x}))\nonumber\\ =&N\frac{1-\mathcal{J}_0\left(\frac{\Gamma_cJ_{\parallel,\mathrm{st}}\tau}{2}\frac{\sin\left(\frac{k_cv_z\tau}{2}\right)}{\frac{k_cv_z\tau}{2}}\right)}{\frac{\Gamma_cJ_{\parallel,\mathrm{st}}\tau}{2}}.\label{Jparresult}
	\end{align}
	Solving this implicit equation for $J_{\parallel,\mathrm{st}}$ and using the result in Eq.~\eqref{Kanalytic} allows us to describe the dipole density in the regular SSR phase. 
	
	\section{Derivation of the phase diffusion model}\label{App:C}
	
	In this section, we integrate Eq.~\eqref{svert} to obtain an analytical result for $J_{\perp}$. This result is used to calculate the linewidth using Eq.~\eqref{Gamma}. Furthermore we use this analytical result to calculate the threshold between the regular SSR and bistable SSR phases. 
	
	Using the Laplace transform on Eq.~\eqref{svert} we obtain
	\begin{align}
		\left[\nu-\mathcal{L}_0\right] L[s_{\perp}]-s_{\perp}({\bf x},0)=\frac{\Gamma_c}{2}\eta s_{z,\mathrm{st}}L[J_{\perp}]+L[\mathcal{S}_{\perp}].\label{sperpdynlaplace}
	\end{align}
	Here we have used the fact that  $s_{z,\mathrm{st}}$ is time independent, and included the definition in Eq.~\eqref{L0}. The initial condition $s_{\perp}({\bf x},0)$ arises from the noisy boundary condition that represents atoms entering the cavity. It is given by
	\begin{align}s_{\perp}({\bf x},0)=W_{\perp}(z-v_z t_0(x),-t_0(x)),
	\end{align}
	where ${t_0(x)=(w+x)/v_x}$.
	Solving now Eq.~\eqref{sperpdynlaplace} for $L[s_{\perp}]$ we find
	\begin{align}
		L[s_{\perp}]=\left[\nu-\mathcal{L}_0\right]^{-1}\left[s_{\perp}({\bf x},0)+\frac{\Gamma_c}{2}\eta s_{z,\mathrm{st}}L[J_{\perp}]+L[\mathcal{S}_{\perp}]\right].
	\end{align}
	Multiplying  by $\eta({\bf x})$ and integrating over the space variable ${\bf x}$, we find an equation for $L[J_{\perp}]$. We solve this equation for $L[J_{\perp}]$ and find the result
	\begin{align}
		L[J_{\perp}]=&\frac{L[J_{W_{\perp}}]+2\frac{1-D_{\perp}(\nu)}{\Gamma_c}L\left[\mathcal{S}_{\perp}\right]}{D_{\perp}(\nu)},\label{LJperp}
	\end{align}
	where 
	\begin{align}
		J_{W_{\perp}}(t)=&\int d{\bf x}\,\eta\left({\bf x}+{\bf v}t\right)W_{\perp}\left(z-v_zt_0(x),-t_0(x)\right)\label{JW}
	\end{align}
	arises from the initial projection noise.  In this derivation, we have used  Eqs.~\eqref{trick1}--\eqref{trick2} and the change of variables given by ${\bf x}\mapsto{\bf x}+{\bf v}t$.
	
	The function $D_{\perp}(\nu)$ is the dispersion relation of the Goldstone mode of the collective dipole, which reads
	\begin{align}
		D_{\perp}(\nu)=&1-\frac{N\Gamma_c}{4w\lambda}\int d{\bf x}\,\int_{0}^\infty dt\,e^{-\nu t}\eta({\bf x}+{\bf v}t)\eta\cos(K).
	\end{align}
	In the regular SSR phase, we can use Eq.~\eqref{DGLK} to rewrite $D_{\perp}(\nu)$ as 
	\begin{align*}
		D_{\perp}(\nu)=&1-\frac{\int_{0}^{\infty} dt e^{-\nu t}\int d{\bf x}\eta\left({\bf x}+{\bf v}t\right){\bf v}\cdot\nabla_{\bf x}s_{\parallel,\mathrm{st}}}{J_{\parallel,\mathrm{st}}},
	\end{align*}
	where
	\begin{align}
		s_{\parallel,\mathrm{st}}=\frac{N}{2w\lambda}\sin(K({\bf x})),\label{sparallel}
	\end{align}
	and $J_{\parallel,\mathrm{st}}=\int d{\bf x}\eta({\bf x})s_{\parallel,\mathrm{st}}$ has been calculated in Eq.~\eqref{Jparresult}.
	
	Applying Gau\ss{}'s theorem and using the fact that the atoms enter in the excited state and that the mode function vanishes at infinity, we get
	\begin{align*}
		D_{\perp}(\nu)=&1+\frac{\int_{0}^{\infty} dte^{-\nu t}\int d{\bf x}\frac{d}{dt}\eta\left({\bf x}+{\bf v}t\right)s_{\parallel,\mathrm{st}}}{J_{\parallel,\mathrm{st}}}.
	\end{align*}
	After another partial integration, we obtain the final form 
	\begin{align}
		D_{\perp}(\nu)=& \nu \frac{\int_{0}^{\infty} e^{-\nu t}dt\int d{\bf x}\eta\left({\bf x}+{\bf v}t\right)s_{\parallel,\mathrm{st}}({\bf x})}{J_{\parallel,\mathrm{st}}}.\label{DispersionGoldstone2}
	\end{align}
	
	The zeros of Eq.~\eqref{DispersionGoldstone2} can be used to describe the dynamics of $J_{\perp}$. In what follows we will assume that $\nu_0=0$ is the solution with the largest real part. With this we can argue that the pole at $\nu=0$ in Eq.~\eqref{LJperp} dictates the long-time behavior of $J_{\perp}$. To describe this long-time behavior we can use the approximation
	\begin{align}
		L[J_{\perp}]\approx&\frac{L[J_{W_{\perp}}]+\frac{2}{\Gamma_c}L\left[\mathcal{S}_{\perp}\right]}{C_0\nu},\label{Jjperpapprox}
	\end{align}
	where 
	\begin{align}
		C_0=\lim_{\nu\to 0}\frac{D_{\perp}(\nu)}{\nu}=\frac{\int_{0}^{\infty} dt\int d{\bf x}\eta\left({\bf x}+{\bf v}t\right)s_{\parallel,\mathrm{st}}({\bf x})}{J_{\parallel,\mathrm{st}}}.\label{C0}
	\end{align}
	By inverting the Laplace transform we find now
	\begin{align}
		J_{\perp}\approx&\frac{\int_{0}^{t} dt'\left[A_1(t')+A_2(t')\right]}{C_0},\label{Jperpdyn}
	\end{align}
	with
	\begin{align}
		A_1(t')=&\int d{\bf x}\,\eta\left({\bf x}+{\bf v}t'\right)W_{\perp}\left(z-v_zt_0(x),-t_0(x)\right),\label{A1}\\
		A_2(t')=&\frac{2\mathcal{S}_{\perp}(t')}{\Gamma_c}.\label{A2}
	\end{align}
	
	Equation~\eqref{Jperpdyn} describes diffusive dynamics perpendicular to the direction of the collective dipole with length $J_{\parallel,\mathrm{st}}$ that results in phase diffusion. Integrating 
	\begin{equation}
		\frac{d\varphi}{dt}\approx \frac{1}{J_{\parallel,\mathrm{ st}}}\frac{dJ_{\perp}}{dt},
	\end{equation}
	we obtain
	\begin{align}
		\Delta\varphi(t)=\varphi(t)-\varphi(0)\approx\frac{\int_{0}^{t} dt'\,\left[A_1(t')+A_2(t')\right]}{C_0J_{\parallel,\mathrm{ st}}}.
	\end{align}
	This result can now be used in Eq.~\eqref{Gamma} to calculate the linewidth.
	
	The diffusive behavior of the phase is a direct result of Eq.~\eqref{Jjperpapprox} where we have assumed that $\nu_0=0$ is a zero of first order of $D_{\perp}(\nu)$. However, it breaks down if $C_0=0$, which indicates that $\nu_0=0$ is a zero of $D_{\perp}(\nu)$ of higher order than first. This can be used to identify the threshold between the regular SSR and bistable SSR phases. We can solve the integrals in Eq.~\eqref{C0} and find
	\begin{align}
		C_0J_{\parallel,\mathrm{ st}}=&\int_{0}^{\infty} dt\int d{\bf x}\eta\left({\bf x}+{\bf v}t\right)s_{\parallel,\mathrm{st}}=\cos\left(\frac{k_cv_z\tau}{2}\right)R,
	\end{align}
	with 
	\begin{align}
		R=2N\int_{0}^1 du\frac{\sin\left(\frac{k_cv_z\tau\left[1-u\right]}{2}\right)\mathcal{J}_1\left(\frac{\Gamma_cJ_{\parallel,\mathrm{st}}\sin\left(\frac{k_cv_z\tau u}{2}\right)}{k_cv_z}\right)}{k_cv_z},
	\end{align}
	where $\mathcal{J}_n$ denotes the Bessel function of order $n$. For this expression we have used the analytical result of $K$ given by Eq.~\eqref{Kanalytic}. For $k_cv_z\tau=\pi$, we obtain $\cos(\pi/2)=0$, hence $C_0=0$, and the phase diffusion argument breaks down. Therefore $k_cv_z\tau=\pi$ is the threshold between regular SSR and bistable SSR.
	
	\section{Superdiffusive behavior at the threshold}\label{App:D}
	In this section we will derive the result in Eq.~\eqref{t^3} using the phase diffusion argument. At the threshold we find that $\nu=0$ is a zero of order two of $D_{\perp}(\nu)$. Using this we can approximate
	\begin{align*}
		L[J_{\perp}]\approx&\frac{L[J_{W_{\perp}}]+\frac{2}{\Gamma_c}L\left[\mathcal{S}_{\perp}\right]}{C_1\nu^2},
	\end{align*}
	where 
	\begin{align}
		C_1=\lim_{\nu\to 0}\frac{D_{\perp}(\nu)}{\nu^2}.\label{C1}
	\end{align}
	This can be used to establish
	\begin{align}
		J_{\perp}\approx&\frac{\int_{0}^{t} dt'\int_{0}^{t'} dt''\left[A_1(t'')+A_2(t'')\right]}{C_1}\label{Jperpdyn2},
	\end{align}
	where we have used Eqs.~\eqref{A1}--\eqref{A2}. Dividing this equation by $J_{\parallel,\mathrm{st}}$ leads to the following equation for the phase
	\begin{align}
		\Delta\varphi(t)\approx\frac{\int_{0}^{t} dt'\int_{0}^{t'} dt''\left[A_1(t'')+A_2(t'')\right]}{C_1J_{\parallel,\mathrm{st}}}.
	\end{align}
	With this one can verify Eq.~\eqref{t^3}.


\begin{thebibliography}{99}
\bibitem{footnote1}
		We have neglected single-particle terms that scale with $\Gamma_c$ under the assumption that ${\Gamma_c\tau\ll1}$. Furthermore, our approach requires that $N\gg1$. In this regime the emission is mostly collective.
\end{thebibliography}
\end{document}